# Hybrid–2D Excitonic Metasurfaces for Complex Amplitude Modulation

Tom Hoekstra[1], Mark L. Brongersma[2], and Jorik van de Groep[1*]

[1] *Van der Waals-Zeeman Institute, Institute of Physics, University of Amsterdam, 1098 XH, Amsterdam, the Netherlands*
[2] *Geballe Laboratory for Advanced Materials, Stanford University, Stanford, CA 94305, USA*

[*] j.vandegroep@uva.nl

**Abstract**
Dynamic control of visible light is crucial for technologies such as holographic displays and adaptive optics. Passive metasurfaces can shape wavefronts at the subwavelength scale and active metasurfaces promise to extend this functionality into the temporal domain. However, existing metasurfaces for dynamic phase manipulation typically cannot deliver phase modulation across a broad range without causing variations in the scattering amplitude. Here, we use an inverse-design pipeline to numerically demonstrate a hybrid-2D excitonic metasurface platform offering independent amplitude and phase control in the visible regime. Harnessing the gate-tunable excitonic response of monolayer $WS_2$ retrieved from experiments, we design a π-phase modulator with a uniform amplitude profile. Adding a second tunable monolayer, we achieve independent control of the amplitude and phase over the full 0–2π phase range, which we leverage for a reconfigurable beam-steering metadevice. Our results demonstrate how hybrid-2D excitonic metasurfaces enable electrically tunable wavefront shaping in the visible regime.

**Keywords:** 2D semiconductor, exciton, inverse design, metasurface, active wavefront shaping

# Introduction

Dynamic wavefront manipulation is crucial for next-generation technologies such as LIDAR[1], holographic displays[2] and adaptive imaging systems[3]. Already, passive metasurfaces enable near-arbitrary shaping of light's constituent properties—amplitude, phase and polarization—by harnessing strong light-matter interactions in planar resonant nanostructure arrays[4]. Building on this foundation, active metasurfaces promise to extend control to the temporal domain by leveraging materials with electrically tunable optical properties[5,6], including ferroelectrics[7,8], conductive oxides[9–14], liquid crystals[15,16], two-dimensional (2D) materials[17–19], and organic polymers[20,21]. Despite these advances, dynamic metasurfaces tend to suffer from undesired cross-modulation of the scattered amplitude and phase, since two independent external controls are required to decouple both parameters[9,22–24]. Additionally, accessing the full 0–2π phase range necessitates a delicate balance between radiative and absorptive losses in the metasurface[25]. Due to these stringent requirements, demonstrations of complete complex amplitude modulation have mostly been limited to the infrared spectral range where materials absorption is naturally lower[9,22,24].

Monolayer transition metal dichalcogenides are promising candidates to extend complex amplitude modulation into the visible regime. These materials exhibit strong exciton resonances that dominate the optical properties due to their large binding energies arising from quantum- and dielectric confinement in the atomically thin crystal[26]. Moreover, these excitons are highly tunable, with free-carrier injection emerging as a particularly attractive mechanism for ultrafast modulation and compatibility with existing electronic circuitry[27–30]. However, limited by their atomic thickness, metasurfaces patterned directly into bare monolayers typically reach efficiencies of only a few percent[31–33]. To enhance the light-matter interaction, monolayers have been placed in the vicinity of metasurfaces to leverage their strong Purcell enhancements[18]. Nevertheless, their performance remains constrained by defects and disorder introduced into the unprotected monolayer during nanofabrication.

Encapsulating the monolayer with hexagonal boron nitride (hBN) provides a robust route to preserve pristine excitonic properties, owing to its atomically smooth surface, wide bandgap, high dielectric strength, and chemical stability[34–36]. The merits of encapsulation are exemplified by experimental demonstrations of planar heterostructures achieving near-unity reflection modulation[37,38] and active beam steering[39,40], as well as full complex amplitude modulation in simulation[24]. Yet, the latter was only achieved by artificially eliminating the non-radiative losses that typically hamper excitonic systems. In practice, these unwanted decay channels can be mitigated by hybridizing planar heterostructures with metasurfaces to leverage strong coupling between excitons and a photonic mode[41]. While this strategy has already delivered record-high excitonic amplitude modulation at room temperature[19], independent phase control remains an important open challenge.

Here, we numerically demonstrate a hybrid-2D metasurface platform for independent amplitude and phase modulation in the visible regime. The high dimensionality of the design space and the intricate interdependence of geometrical and material parameters limit the utility of conventional forward-design approaches. We therefore develop an inverse-design pipeline that employs rigorous coupled-wave analysis (RCWA) to optimize metasurface designs for π-phase modulation and full complex amplitude modulation. We further validate the platform's applicability by demonstrating a three-level programmable beam-steering metasurface. Crucially, all simulations are grounded in experimentally measured material properties and are interpreted using temporal coupled-mode theory (TCMT) to elucidate the underlying physics and establish the generality of our approach.

## Results

### Hybrid-2D phase modulator

Figure 1a shows an active hybrid-2D metasurface for free-space phase modulation, where the gate voltage ($V_G$) shifts the phase ($\varphi$) of a reflected beam while maintaining a constant amplitude. The device combines the tunability of excitons in monolayer tungsten disulfide ($WS_2$) and the field enhancement offered by a non-local dielectric metasurface (Fig. 1b). The exact metasurface architecture is obtained using a bespoke inverse design optimization routine that combines Bayesian optimization and an evolutionary algorithm (Supplementary Note 1) to maximize phase modulation performance under realistic fabrication constraints.

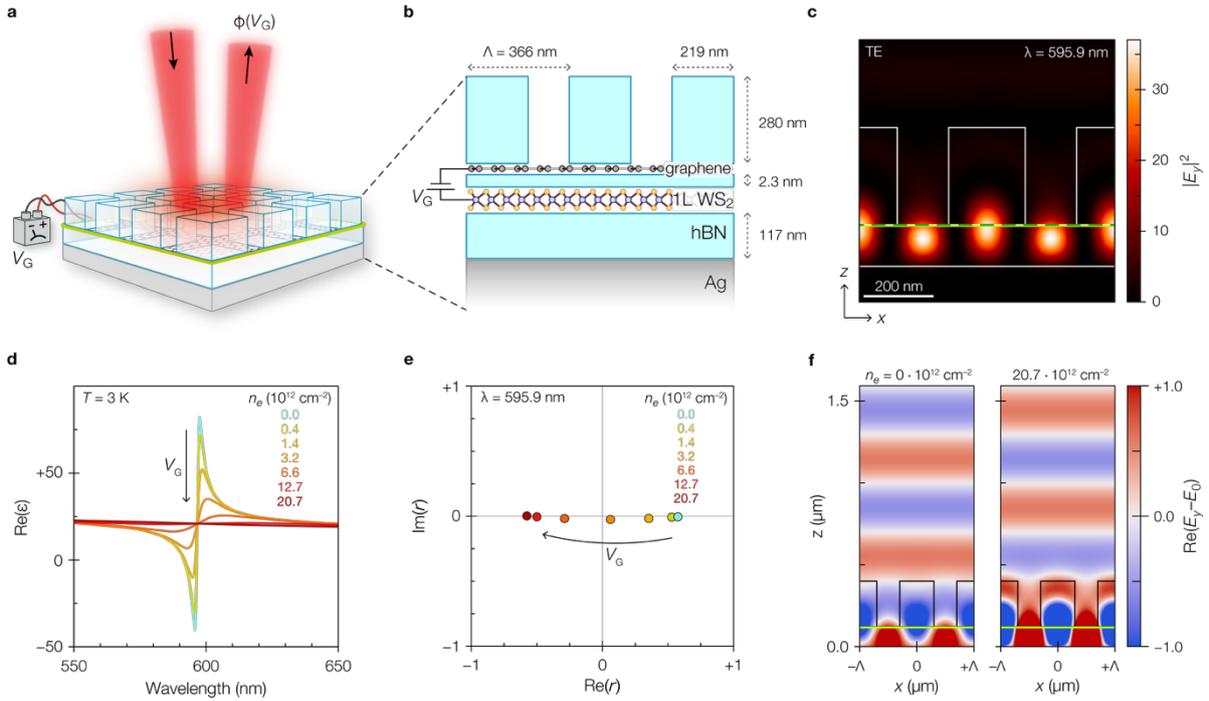

**Figure 1: Hybrid-2D phase modulator. (a)** Conceptual illustration of the hybrid-2D metasurface for free-space phase modulation, in which the gate voltage ($V_G$) controls the phase ($\varphi$) of the reflected light. **(b)** Cross-sectional view of the hybrid-2D metasurface design (not to scale), showing the hBN subwavelength grating, graphene gate electrode, monolayer (1L) $WS_2$ encapsulated by hBN and placed on the Ag mirror. The grating period ($\Lambda$) and other geometrical parameters are indicated. **(c)** Electric field intensity ($|E_y|^2$) enhancement in absence of excitons, for a transverse electric (TE) polarized normal-incident wave at wavelength $\lambda = 595.9$ nm. The positions of 1L $WS_2$ (light green, solid) and graphene (dark green, dashed) are indicated. **(d)** Modeled permittivity of monolayer $WS_2$ as function of electron density ($n_e$, indicated by color) at temperature $T = 3$ K, and **(e)** corresponding complex reflection coefficients of the metasurface at $\lambda = 595.9$ nm. **(f)** Corresponding electric field distributions of the reflected light ($\text{Re}(E_y - E_0)$) for intrinsic ($n_e = 0.0 \times 10^{12}$ cm$^{-2}$, left) and highly n-doped ($n_e = 20.7 \times 10^{12}$ cm$^{-2}$, right) 1L $WS_2$ yielding the same amplitude but opposite phase.

The designed modulator combines several key elements. First, a monolayer $WS_2$ is encapsulated between two multi-layer slabs of hBN, which is known to reduce the linewidth of the exciton resonance by screening it from the dielectric environment[42,43]. The entire stack is situated on a silver (Ag) back-reflector to effectively trap light in the heterostructure cavity[34] and redirect it into the reflection channel. The heterostructure is then functionalized with a planar array of hBN subwavelength scatterers spaced with period $\Lambda = 366$ nm, enabling polarization-independent normal-incident resonant excitation of a guided mode resonance (GMR) which provides a 36-fold enhancement of the incident field intensity (Fig. 1c, transverse magnetic (TM) polarization shown in Supplementary Fig. S1). A single layer of graphene, situated beneath the grating, acts as an etch stop during dry etching of the hBN nanostructures[44] and serves as the gate electrode for modulation of the electron density ($n_e$) in the $WS_2$ monolayer. The thickness of the hBN gate dielectric is 7 atomic layers thick (2.33 nm) to provide a sufficiently high breakdown voltage[35,36] (Supplementary Fig. S2) to enable significant amplitude and phase modulation at low-voltages (0–5 V). By inducing n-type doping through electrostatic gating, the

A-exciton resonance ($\lambda_A$ = 596.7 nm) can be actively and reversibly quenched, thereby modulating the material's permittivity[27–30] (Fig. 1d).

We model the permittivity of monolayer WS$_2$ using a phenomenological Lorentz-oscillator model that effectiely captures the 2D sheet response within the 3D permittivity as[27,33,45–47]

$$\varepsilon(\omega) = \varepsilon_{bg} + \frac{c}{d}\sum_j \frac{\gamma_{j,r}}{\omega_j^2 - \omega^2 - i\omega\gamma_j}$$

Here, $\omega$ denotes the frequency, $\varepsilon_{bg}$ is the background permittivity, $c$ is the speed of light, $d$ = 6.18 Å is the thickness of monolayer WS$_2$,[48] and the $j^{th}$ exciton is characterized by its resonance frequency $\omega_j$ and decay rate $\gamma_j = \gamma_{j,r} + \gamma_{j,nr}$, which is the sum of the radiative and non-radiative components, respectively (where we absorb other broadening mechanisms such as pure dephasing[49,50] in the effective linewidth). By fitting reflectance measurements performed at cryogenic temperature ($T$ = 3 K, Supplementary Fig. S3), we parametrize the A-exciton with resonance energy $E_A = \hbar\omega_A$ = 2.078 eV, $\gamma_{A,r}$ = 2.6 meV, and $\gamma_{A,nr}$ = 3.9 meV. Note that these exciton decay rates are not universal material constants, but depend on the adopted permittivity convention[50], as well as on the sample configuration, for instance through the local density of optical states and local Coulomb screening[51,52]. These values correspond to an excitonic quantum yield of $\gamma_{A,r} / (\gamma_{A,r} + \gamma_{A,nr})$ = 0.4, which is realistically achievable in high-quality monolayer WS$_2$. In this context, atomically thin MoSe$_2$ appears to be a particularly promising alternative, as recent measurements identified it as the most efficient emitter within the transition metal dichalcogenide family[53]. Although near-unity quantum yields have been reported for these materials in isolated cases[37,54], such performance is not routinely observed under standard laboratory conditions.

The electron density as function of gate bias is modeled with a parallel-plate capacitor model corrected for quantum capacitance effects[55,56] (Methods). From this, we phenomenologically capture the doping-dependent quenching of the A-exciton resonance by letting $\gamma_{A,nr}$ increase quadratically with $n_e$ (Supplementary Fig. S2), in accordance with previous experimental work[19,57]. We typically do not resolve a distinct trion resonance in coherent reflection measurements, which may be attributed to its comparatively low oscillator strength and large non-radiative decay rate[57], and we therefore treat trion formation as an additional exciton loss channel. We additionally note that the tunable range of $\gamma_{A,nr}$ is extrapolated to high carrier densities and assumes the possibility to completely quench the resonance. If this limit cannot be reached experimentally, the inverse-design pipeline can readily re-optimize the metasurface geometry for any realistic exciton amplitude. Under this model, as the exciton is gradually quenched, the complex reflection coefficient at $\lambda$ = 595.9 nm traces a path from $\varphi$ = 0 rad to $\varphi$ = $\pi$ rad (Fig. 1e). Thus, when the monolayer is switched between intrinsic and strong n-type doping ($n_e$ = 20.7×10$^{12}$ cm$^{-2}$), the device flips the phase of a reflected beam while maintaining a constant amplitude (Fig. 1f). Although this doping level exceeds what is typically reported in electrostatic gating studies, ultrathin hBN has been shown to sustain high out-of-plane dielectric fields[35], indicating that this carrier-density range is not fundamentally ruled out by dielectric breakdown.

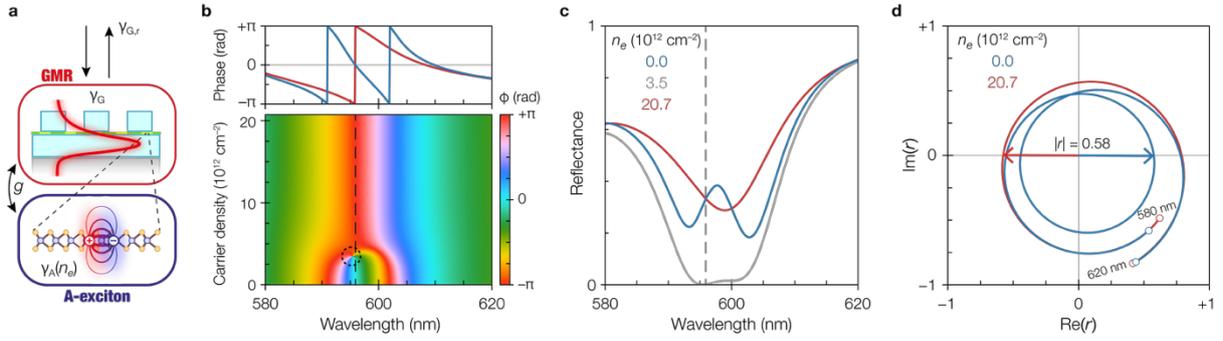

**Figure 2: Operation near critical coupling enables π-phase modulation. (a)** Schematic of the one-port coupled-resonator system. The guided mode resonance (GMR) has intrinsic loss rate $\gamma_G$ and facilitates free-space coupling with radiative rate $\gamma_{G,r}$. The A-exciton is coupled to the GMR with strength $g$, and its decay rate $\gamma_A$ is modulated by gate-induced changes in carrier density $n_e$. **(b)** Top: reflection-phase spectra at intrinsic (blue) and strong n-doping (red). Bottom: phase ($\varphi$) as function of carrier density, with the operating wavelength indicated (dashed line). The dashed black circle indicates the phase singularity arising from critical coupling. **(c)** Reflectance spectra at different carrier concentrations showing the evolution from under-coupled at intrinsic doping (blue) to critically coupled at mild n-doping (gray) to over-coupled at strong n-doping (red). Dashed line indicates the operating wavelength. **(d)** Phasor diagram of the complex reflection coefficient, $r$, at intrinsic (blue) and strong n-doping (red). Vectors indicate $r$ at the operating wavelength.

Figure 2a schematically depicts how light interacts with guided-mode and A-exciton resonances in our metasurface. Free-space radiation couples to the GMR with radiative rate $\gamma_{G,r}$ when the resonance condition is satisfied (matching of the in-plane wavevector by addition of grating momentum $G = 2\pi/\Lambda$). In turn, the GMR can coherently excite excitons through optical coupling, with coupling strength $g$ governed by the field overlap between both resonances. From fitting of the reflectance in absence of excitons (achieved by setting $\gamma_{A,r} = 0$ meV to remove the A-exciton contribution from the permittivity of WS$_2$), we determine the radiative coupling rate and the non-radiative loss rate of the GMR, $\gamma_{G,r} = 55.0$ meV and $\gamma_{G,nr} = 13.8$ meV, respectively. Since $\gamma_{G,r} \gg \gamma_{A,r}$, we consider direct coupling between excitons and free-space to be negligible, making this effectively a one-port system.

To achieve the π phase flip, it is necessary to operate around the critical coupling point ($|r| = 0$), where the phase is undefined and thus singular (Fig. 2b). In a one-port system, this condition occurs when the metasurface's radiative coupling rate matches the overall loss rate, which is set by dissipation in the Ag, graphene, and monolayer WS$_2$[58,59]. In this spectral range and at these carrier densities, the optical response of graphene is expected to vary only weakly with gating[60,61]. We therefore separate carrier-density-independent background losses suffered in Ag, graphene and WS$_2$ (from inter-band transitions), captured by $\gamma_{G,nr}$, from the electrically tunable excitonic absorption in monolayer WS$_2$, captured by $\gamma_{A,nr}(n_e)$. In our device, critical coupling is reached when the monolayer is at mild doping, $n_e = 3.5 \times 10^{12}$ cm$^{-2}$ (Fig. 2c). At strong n-doping, the exciton resonance is quenched and the bare GMR produces a single reflectance minimum. On the other hand, at intrinsic doping, the exciton hybridizes with the GMR, leading to an avoided crossing in the dispersion that appears as two distinct reflectance minima. For both regimes, we plot the spectral evolution of the reflection amplitude and phase in a phasor diagram (Fig. 2d). In the n-doped case, the reflection coefficient traces a single loop in the complex plane (single photonic resonance), whereas at charge neutrality a double loop is traced due to the additional presence of the A-exciton (material resonance). At the design wavelength $\lambda = 595.9$ nm, perfect π-phase modulation is achieved with a constant amplitude of $|r| = 0.58$.

Next, we probe the fundamental limits of a π-phase modulator by performing a TCMT analysis of a one-port two-resonator system (Supplementary Fig. S4 and Supplementary Note 2). We estimate $g \leq 20$ meV as a reasonable physical bound based on previous work[19], which we plug into our model together with the stated radiative and non-radiative rates. Using our inverse design framework, we find an upper bound on the reflection amplitude that allows π-phase modulation of $|r|_{max} = 0.62$, which highlights that our modulator design ($|r| = 0.58$) operates close to this fundamental limit. The TCMT model additionally allows us to quantify how changes in the metasurface's fundamental rates affect the achievable phase modulation depth (Supplementary Fig. S5). For instance, increasing the radiative coupling rate of the GMR could be achieved by raising the index contrast of the subwavelength grating through duty cycle adjustments or by using a higher-index material. Reducing absorption losses may be achieved by minimizing losses in the electrical contacts and in the back-reflector (*e.g.* using a dielectric Bragg mirror). Moreover, the excitonic quantum yield could be further optimized by improving the material quality of $WS_2$ or via strain engineering[54]. Overall, we find that cutting the non-radiative losses provides the largest gain, with the maximum modulation amplitude increasing to $|r|_{max}$ = 0.81. Thus, eliminating parasitic absorption channels in the metasurface is the most fruitful avenue to further enhance the π-phase modulation depth.

Altogether, these results demonstrate a general strategy for achieving π-phase modulation by sweeping through critical coupling using a single control parameter[10]. Intriguingly, this concept can be extended to gain independent control over amplitude and phase—and thus achieve complete complex amplitude modulation—by introducing a second control parameter.

**Complex amplitude modulation using two independent control voltages**
The two-control-parameter approach has previously been demonstrated as an effective route to full complex-reflection control in the infrared spectral domain[9,22,24]. Here, we leverage the remarkable tunability of excitons in our hybrid-2D metasurface platform to extend this mechanism into the visible regime. To achieve this, we add a second monolayer at a different position in the heterostructure stack (Fig. 3a), which is independently addressable by applying a voltage ($V_B$) from the Ag back-reflector, while the original monolayer remains electrically connected to the graphene gate ($V_T$). Both monolayers are electrically insulated from each other by an extra multi-layer hBN slab. To enhance the modulation depth with this architecture, we choose to relax the maximum voltage constraint to $V_{max}$ = 10 V. Again, we optimize the metasurface geometry using our inverse design framework, updating the figure-of-merit to maximize the amplitude $|r|$ at which complex amplitude modulation is achievable over the full 0–2π phase range.

In the optimized design, both monolayers are separated by 102 nm of hBN and therefore experience different field overlaps with the GMR (Fig. 3b, TM polarization shown in Supplementary Fig. S1). The excitons in the top and bottom monolayers therefore couple to the GMR with unequal strengths, breaking the symmetry between both resonances (Supplementary Note 2 and Supplementary Fig. S6). Consequently, the complex $r(n_t, n_b)$ is uniquely affected by gate-modulation of the top and bottom carrier densities ($n_t$ and $n_b$, respectively), with the extrema of $n_t$ (0 to $25\times10^{12}$ cm$^{-2}$) and $n_b$ (0 to $6.8\times10^{12}$ cm$^{-2}$) roughly defining the perimeter of the tunable range (Fig. 3c).

To access the full 0–2π phase modulation range, the metasurface is designed to center $r$ around the origin of the complex plane and thus operate in the vicinity of critical coupling (Fig. 3d,e and Supplementary Fig. S7). Whereas with a single control parameter it is only possible to trace a straight trajectory through the phase singularity, the two-control-parameter approach enables phase-only modulation by tracing a constant-reflectance path around a phase singularity (Fig. 3f). By judiciously selecting combinations of ($n_t$, $n_b$), we trace the largest constant-amplitude contour that sweeps the full 0–2π phase range at $|r|$ = 0.13. We note that the maximum amplitude tunability extends beyond this cut-off with a limited phase range, which can be advantageous for applications including beam steering[62]. Notably, the same mechanism remains functional at room temperature, albeit it with a reduced constant-amplitude of $|r|$ = 0.06 owing to the larger excitonic non-radiative decay rate (Supplementary Fig. S8).

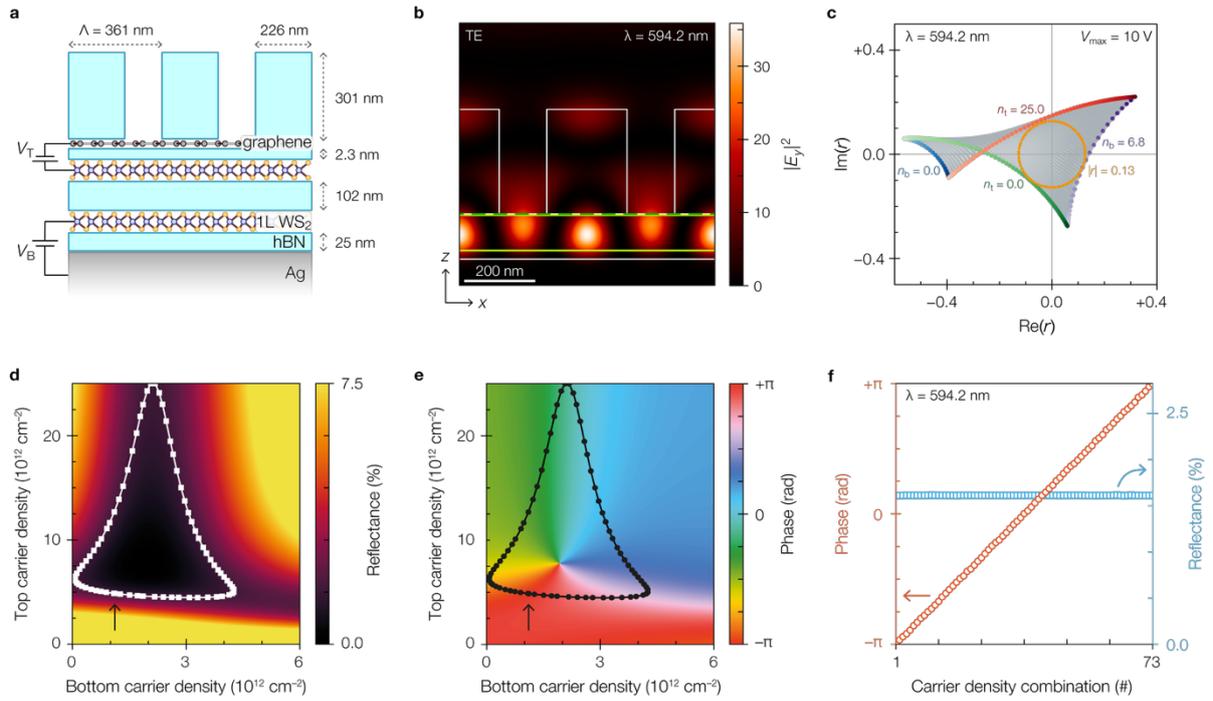

**Figure 3: Complex amplitude modulation with two independent control voltages. (a)** Cross-sectional view of the $2\pi$ phase modulator (not to scale). The $WS_2$ monolayers (1L) are independently addressable via the top ($V_T$) and bottom ($V_B$) gate voltages. The grating period ($\Lambda$) and other geometrical parameters are indicated. **(b)** Electric field intensity ($|E_y|^2$) in absence of excitons, for a transverse electric (TE) polarized normal-incident wave at wavelength $\lambda = 594.2$ nm. The positions of 1L $WS_2$ (light green, solid) and graphene (dark green, dashed) are indicated. **(c)** The reflection coefficients, $r$, which result from all possible combinations of carrier densities in the top ($n_t$) and bottom ($n_b$) monolayers (the extrema are color-marked, units: $10^{12}$ cm$^{-2}$). The metasurface achieves complete complex amplitude modulation up to $|r| = 0.13$ (orange circle). **(d)** Reflectance map and **(e)** reflection-phase map as function of top and bottom carrier densities. The carrier densities corresponding to $|r| = 0.13$ are indicated. **(f)** Continuous phase modulation from $-\pi$ to $\pi$ at a constant reflection amplitude ($|r|^2 = 0.13^2$). The carrier density combinations correspond to those shown in (d) and (e), where combination #1 is indicated with an arrow.

Again, we use TCMT to assess whether the observed reflection amplitude approaches the theoretical maximum (Supplementary Fig. S9). First, we fit the reflectance without excitons to obtain the GMR loss rates, $\gamma_{G,r} = 55.0$ meV and $\gamma_{G,nr} = 18.6$ meV. Notably, the non-radiative rate is slightly higher than before due to the addition of another $WS_2$ monolayer in the heterostructure cavity. Even without excitons, the monolayers exhibit sub-bandgap absorption from impurities, edge states and defects[63], which increases the overall absorption losses. We consider a one-port resonator with three coupled oscillators (GMR and two exciton resonances) and optimize the exciton-photon coupling strengths, the metasurface's resonance wavelength, and the operating wavelength. The resulting TCMT system achieves full complex amplitude modulation at $|r| = 0.15$, indicating that our inverse-designed metasurface operates close the fundamental limits of a reflection-phase modulator with the stated internal and external loss rates. Naturally, the modulation depth is affected by modifying the fundamental rates that govern the metasurface's optical response (Supplementary Fig. S10). We again observe the strongest enhancement when absorption losses in the metasurface are eliminated ($\gamma_{G,nr} = 0$ meV), increasing the full-$2\pi$ amplitude to $|r| = 0.22$. Beyond this, optimizing the radiative coupling rate, improving the excitonic quantum yield, or extending the modulation range of excitons in the bottom monolayer each yield only modest additional gains.

**Three-level programmable beam steering metasurface**

To demonstrate the utility of complex amplitude modulation using our hybrid-2D metasurface platform, we design a three-level programmable beam steering device. Building on the double-gated design outlined above, we create a supercell with period $\Lambda = 1.08$ μm consisting of three geometrically identical unit cells (Fig. 4a). The $WS_2$ monolayers are severed by carving 11-nm-wide slits along the $y$-axis to define individually addressable strips in the top and bottom monolayers (six gates in total). This segmentation lets us program the complex reflection coefficient of each unit cell (under TM illumination) and enables sculpting of the reflected wavefront across the supercell.

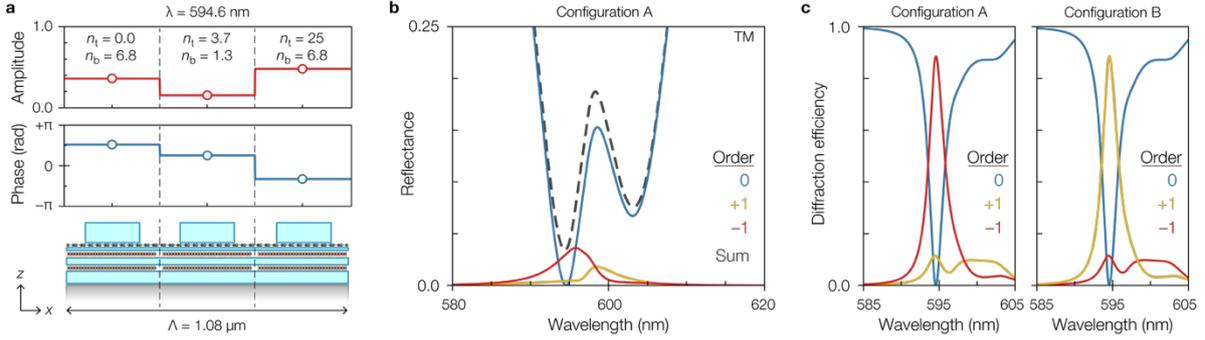

**Figure 4: Beam steering with a three-level programmable metasurface. (a)** Bottom: three-level programmable metasurface with supercell period $\Lambda$, designed for beam steering to the –1 order (configuration A). The phase (middle) and amplitude (top) of each level is programmed independently by tuning the top ($n_t$) and bottom ($n_b$) carrier densities to the values indicated (units: $10^{12}$ cm$^{-2}$) **(b)** Reflected power into the –1 (red), 0 (blue), and +1 (yellow) orders for configuration A. The total reflectance is indicated by the dashed line. **(c)** Left: corresponding diffraction efficiency reveals 88.5% of reflected power is redirected into the –1 order. Right: steering light into the +1 order is achieved by programming the mirror-image of configuration A.

As a baseline, we design an intuitive forward-designed beam steering metasurface—a phased array with a flat amplitude profile and a constant phase gradient, *i.e.*, a blazed grating (Supplementary Fig. S11). Picking three carrier density combinations from Fig. 3f with relative phases spaced $2\pi/3$ rad apart, we design the phase ramp to steer light into the –1 order (at an angle of $\theta = -33.3°$). We find that this phased array reaches a maximum diffraction efficiency of 77.2% with only $R = 0.3\%$ reflected power at $\lambda = 594.8$ nm.

To improve on the forward-designed device, we simultaneously engineer both amplitude and phase into a less intuitive configuration, which has been shown to enhance performance[62,64]. Again, we utilize our inverse design routine, keeping all layer thicknesses from the preceding design but varying the grating-element width to enlarge the tuning range of the reflection amplitude (Supplementary Fig. S12). For configuration A, we jointly optimize the element width and the carrier densities in the top and bottom monolayers to generate a three-level amplitude and phase pattern that steers light into the –1 order (Fig. 4a). This yields a marked improvement, with the total reflected power at the operating wavelength ($\lambda = 594.6$ nm) rises to $R = 3.4\%$ (Fig. 4b) and the power is re-directed more efficiently, with 88.5% diffracted into the –1 order (Fig. 4c).

The same programmable architecture can be reconfigured to enable other functionalities. Reprogramming the carrier densities to achieve the spatial mirror image of the amplitude-phase profile (configuration B), the diffraction pattern is inverted, swapping the +1 and –1 order (Fig. 4c). Naturally, increasing the number of independently tunable elements within the supercell would transform the system from a coarse three-level array into a fully programmable spatial light modulator. This comes with the practical limitation of our current design's reliance on a non-local resonance, which imposes a larger unit-cell footprint than highly localized resonances would. Yet, a hybrid approach that leverages both local and nonlocal resonances within the same structure may mitigate this constraint while enabling multifunctional or multi-wavelength operation[65,66]. Overall, these results demonstrate the flexibility of our hybrid-2D platform for electrically reconfigurable metasurfaces.

## Discussion and conclusions

In this work, we establish hybrid-2D metasurfaces as a versatile platform for active wavefront shaping, enabled by the electrical tunability of excitons in Van der Waals heterostructures combined with the field enhancement of a non-local resonance. Inverse design is used throughout this work to maximize the performance within the physical constraints. With a single gate-tunable monolayer WS$_2$, sweeping through critical coupling enables a $\pi$ phase flip at constant amplitude. The metasurface reaches an amplitude of $|r| = 0.58$ which closely tracks the upper bound of $|r| = 0.62$ retrieved from TCMT. Incorporating a second, independently gated monolayer WS$_2$ extends the phase coverage to 0–2$\pi$ and decouples the amplitude to realize full complex amplitude modulation, which the optimized design achieves up to $|r| = 0.13$, agreeing well with the TCMT ceiling of $|r| = 0.15$. Finally, we demonstrate a three-level programmable metasurface for beam steering. The device re-directs light with 88.5% efficiency into the ±1 order at $\theta = \pm33.3°$, with the $0^{th}$ order entirely suppressed, underscoring the applicability of our hybrid-2D platform.

The modest amplitudes associated with full complex-amplitude control reflect the need to operate near critical coupling to access the entire 0–2$\pi$ phase range—a trend seen across other metasurface modulators. In an experimental realization[9], a conductive oxide plasmonic metasurface is capped at $|r| \approx 0.08$, while graphene plasmonic metamolecules reach $|r| \approx 0.19$ in simulation[22]. Yet, both approaches are restricted to polarization-dependent operation in the infrared spectral region. More recently, these limitations were overcome with an excitonic 2D materials heterostructure that enables polarization-independent complex-amplitude control just beyond the visible regime[24]. However, this work relies on unity excitonic quantum yield—a condition rarely observed under standard laboratory conditions. Against this backdrop, achieving visible-wavelength, polarization-insensitive complex amplitude modulation with realistic material parameters represents a meaningful advance for metasurface modulators.

As shown throughout this work, the efficiency of our modulator is ultimately constrained by the radiative and non-radiative channels within the metasurface. Improving the demonstrated design therefore hinges on reducing absorption or enhancing radiative coupling. The former may be achieved by minimizing Ohmic losses in the gate electrodes and suppressing sub-bandgap absorption in the excitonic materials by eliminating impurities and defects. The latter could be accomplished by replacing hBN in the grating layer with a higher-index, low-loss material such as TiO$_2$ or Si. In practice, an important consideration for this platform is fabrication imperfections such as interlayer contamination and strain in the heterostructure or sidewall roughness in the grating, which may lead to spectral shifts or reduced modulation depth. However, by now, it is increasingly possible to mitigate these adverse effects through refinements in stacking, cleaning, and dry etching of 2D materials and their heterostructures[41,67–69], which minimize spectral broadening due to spatial inhomogeneities.

From a technological perspective, the performance and scalability of 2D materials remains a key challenge. Yet, substantial progress is being made through advances in synthesis[70], large-area exfoliation[71,72], transfer techniques[67] and robotic assembly[73], bringing wafer-scale integration increasingly within reach. Although the modulation bandwidth and energy were not investigated here, comparable device geometries have already demonstrated gigahertz operation[39,74,75], with switching energies expected to lie in the low-femtojoule range[75,76]. Altogether, our results establish inverse-designed hybrids combining non-local metasurfaces and excitonic 2D materials heterostructures as a promising platform for dynamically reconfigurable metasurfaces.

**Supporting Information**
Methods for sample fabrication, cryogenic spectroscopy experiments, inverse design, optical modeling; additional details on inverse design protocol and coupled mode theory; including Figures S1-S12 with electric field profiles, reflectance fitting, and carrier density calculations. (PDF).

**Acknowledgements**
This work was funded by a Vidi grant (VI.Vidi.203.027) from the Dutch National Science Foundation (NWO). JvdG is also supported by an European Research Council Starting Grant under grant agreement No. 101116984. We express our gratitude to Sander A. Mann for insightful discussions about the TCMT model.

**Data availability**
A full replication package including all data and scripts will be made openly available upon publication.

**Conflict of interest**
The authors declare no competing interests.

**Contributions**
T.H., J.v.d.G. conceived the concept behind this research. T.H. fabricated the sample, performed the measurements, and performed the RCWA and TCMT simulations with input from J.v.d.G. and M.L.B. All authors contributed to analyzing the results and writing the manuscript.

# References

1. Kim, I. *et al.* Nanophotonics for light detection and ranging technology. *Nat. Nanotechnol.* **16**, 508–524 (2021).
2. Gopakumar, M. *et al.* Full-colour 3D holographic augmented-reality displays with metasurface waveguides. *Nature* **629**, 791–797 (2024).
3. Hampson, K. M. *et al.* Adaptive optics for high-resolution imaging. *Nat. Rev. Methods Primer* **1**, 68 (2021).
4. Kuznetsov, A. I. *et al.* Roadmap for Optical Metasurfaces. *ACS Photonics* **11**, 816–865 (2024).
5. Shaltout, A. M., Shalaev, V. M. & Brongersma, M. L. Spatiotemporal light control with active metasurfaces. *Science* **364**, eaat3100 (2019).
6. Gu, T., Kim, H. J., Rivero-Baleine, C. & Hu, J. Reconfigurable metasurfaces towards commercial success. *Nat. Photonics* **17**, 48–58 (2023).
7. Klopfer, E., Dagli, S., Barton, D. I., Lawrence, M. & Dionne, J. A. High-Quality-Factor Silicon-on-Lithium Niobate Metasurfaces for Electro-optically Reconfigurable Wavefront Shaping. *Nano Lett.* **22**, 1703–1709 (2022).
8. Dagli, S. *et al.* GHz-Speed Wavefront Shaping Metasurface Modulators Enabled by Resonant Electro-Optic Nanoantennas. *Adv. Mater.* **n/a**, e06790 (2025).
9. Park, J. *et al.* All-solid-state spatial light modulator with independent phase and amplitude control for three-dimensional LiDAR applications. *Nat. Nanotechnol.* **16**, 69–76 (2021).
10. Park, J., Kang, J.-H., Kim, S. J., Liu, X. & Brongersma, M. L. Dynamic Reflection Phase and Polarization Control in Metasurfaces. *Nano Lett.* **17**, 407–413 (2017).
11. Shirmanesh, G. K., Sokhoyan, R., Wu, P. C. & Atwater, H. A. Electro-optically Tunable Multifunctional Metasurfaces. *ACS Nano* **14**, 6912–6920 (2020).
12. Huang, Y.-W. *et al.* Gate-Tunable Conducting Oxide Metasurfaces. *Nano Lett.* **16**, 5319–5325 (2016).
13. Kafaie Shirmanesh, G., Sokhoyan, R., Pala, R. A. & Atwater, H. A. Dual-Gated Active Metasurface at 1550 nm with Wide (>300°) Phase Tunability. *Nano Lett.* **18**, 2957–2963 (2018).
14. Sisler, J. *et al.* Electrically tunable space–time metasurfaces at optical frequencies. *Nat. Nanotechnol.* **19**, 1491–1498 (2024).
15. Komar, A. *et al.* Dynamic Beam Switching by Liquid Crystal Tunable Dielectric Metasurfaces. *ACS Photonics* **5**, 1742–1748 (2018).

16. Hu, Y. *et al.* Electrically Tunable Multifunctional Polarization-Dependent Metasurfaces Integrated with Liquid Crystals in the Visible Region. *Nano Lett.* **21**, 4554–4562 (2021).

17. Lynch, J. *et al.* Full 2π phase modulation using exciton-polaritons in a two-dimensional superlattice. *Device* **3**, (2025).

18. Li, Q. *et al.* A Purcell-enabled monolayer semiconductor free-space optical modulator. *Nat. Photonics* **17**, 897–903 (2023).

19. Hoekstra, T. & van de Groep, J. Electrically tunable strong coupling in a hybrid-2D excitonic metasurface for optical modulation. *Light Sci. Appl.* **15**, 28 (2026).

20. Doshi, S. *et al.* Electrochemically mutable soft metasurfaces. *Nat. Mater.* **24**, 205–211 (2025).

21. Benea-Chelmus, I. C. *et al.* Gigahertz free-space electro-optic modulators based on Mie resonances. *Nat. Commun.* **13**, 3170 (2022).

22. Han, S. *et al.* Complete Complex Amplitude Modulation with Electronically Tunable Graphene Plasmonic Metamolecules. *ACS Nano* **14**, 1166–1175 (2020).

23. Overvig, A. C. *et al.* Dielectric metasurfaces for complete and independent control of the optical amplitude and phase. *Light Sci. Appl.* **8**, 92 (2019).

24. Li, M., Michaeli, L. & Atwater, H. A. Electrically Tunable Topological Singularities in Excitonic Two-Dimensional Heterostructures for Wavefront Manipulation. *ACS Photonics* **11**, 3554–3562 (2024).

25. Kim, J. Y. *et al.* Full 2π tunable phase modulation using avoided crossing of resonances. *Nat. Commun.* **13**, 2103 (2022).

26. Chernikov, A. *et al.* Exciton Binding Energy and Nonhydrogenic Rydberg Series in Monolayer $WS_2$. *Phys. Rev. Lett.* **113**, 076802 (2014).

27. Li, M., Biswas, S., Hail, C. U. & Atwater, H. A. Refractive Index Modulation in Monolayer Molybdenum Diselenide. *Nano Lett.* **21**, 7602–7608 (2021).

28. Yu, Y. *et al.* Giant Gating Tunability of Optical Refractive Index in Transition Metal Dichalcogenide Monolayers. *Nano Lett.* **17**, 3613–3618 (2017).

29. Chernikov, A. *et al.* Electrical Tuning of Exciton Binding Energies in Monolayer $WS_2$. *Phys. Rev. Lett.* **115**, 126802 (2015).

30. Ross, J. S. *et al.* Electrical control of neutral and charged excitons in a monolayer semiconductor. *Nat. Commun.* **4**, 1474 (2013).

31. van de Groep, J. *et al.* Exciton resonance tuning of an atomically thin lens. *Nat. Photonics* **14**, 426–430 (2020).

32. Guarneri, L. *et al.* Dynamic Excitonic Beam Switching with Atomically-Thin Binary Blazed Gratings. *Adv. Opt. Mater.* 2403257 (2025) doi:10.1002/adom.202403257.

33. Guarneri, L. *et al.* Temperature-Dependent Excitonic Light Manipulation with Atomically Thin Optical Elements. *Nano Lett.* **24**, 6240–6246 (2024).

34. Epstein, I. *et al.* Near-unity light absorption in a monolayer $WS_2$ van der Waals heterostructure cavity. *Nano Lett.* **20**, 3545–3552 (2020).

35. Ranjan, A. *et al.* Dielectric Breakdown in Single-Crystal Hexagonal Boron Nitride. *ACS Appl. Electron. Mater.* **3**, 3547–3554 (2021).

36. Hattori, Y., Taniguchi, T., Watanabe, K. & Nagashio, K. Anisotropic Dielectric Breakdown Strength of Single Crystal Hexagonal Boron Nitride. *ACS Appl. Mater. Interfaces* **8**, 27877–27884 (2016).

37. Back, P., Zeytinoglu, S., Ijaz, A., Kroner, M. & Imamoğlu, A. Realization of an Electrically Tunable Narrow-Bandwidth Atomically Thin Mirror Using Monolayer $MoSe_2$. *Phys. Rev. Lett.* **120**, 03740 (2018).

38. Scuri, G. *et al.* Large Excitonic Reflectivity of Monolayer $MoSe_2$ Encapsulated in Hexagonal Boron Nitride. *Phys. Rev. Lett.* **120**, 037402 (2018).

39. Andersen, T. I. *et al.* Beam steering at the nanosecond time scale with an atomically thin reflector. *Nat. Commun.* **13**, 3431 (2022).

40. Li, M., Hail, C. U., Biswas, S. & Atwater, H. A. Excitonic Beam Steering in an Active van der Waals Metasurface. *Nano Lett.* **23**, 2771–2777 (2023).

41. Sortino, L. *et al.* Atomic-layer assembly of ultrathin optical cavities in van der Waals heterostructure metasurfaces. *Nat. Photonics* **19**, 825–832 (2025).

42. Moody, G. *et al.* Intrinsic homogeneous linewidth and broadening mechanisms of excitons in monolayer transition metal dichalcogenides. *Nat. Commun.* **6**, 8315 (2015).

43. Cadiz, F. *et al.* Excitonic linewidth approaching the homogeneous limit in $MoS_2$-based van der Waals heterostructures. *Phys. Rev. X* **7**, 021026 (2017).

44. Danielsen, D. R. *et al.* Super-Resolution Nanolithography of Two-Dimensional Materials by Anisotropic Etching. *ACS Appl. Mater. Interfaces* **13**, 41886–41894 (2021).

45. Li, Y. *et al.* Measurement of the optical dielectric function of monolayer transition-metal dichalcogenides: $MoS_2$, $MoSe_2$, $WS_2$, and $WSe_2$. *Phys. Rev. B* **90**, 205422 (2014).

46. Hsu, C. *et al.* Thickness-Dependent Refractive Index of 1L, 2L, and 3L $MoS_2$, $MoSe_2$, $WS_2$, and $WSe_2$. *Adv. Opt. Mater.* **7**, 1900239 (2019).

47. Ivčenko, E. L. *Optical Spectroscopy of Semiconductor Nanostructures*. (Alpha Science International Ltd, Harrow, UK, 2005).

48. Wilson, J. A. & Yoffe, A. D. The transition metal dichalcogenides discussion and interpretation of the observed optical, electrical and structural properties. *Adv. Phys.* **18**, 193–335 (1969).

49. Van de Groep, J., Li, Q., Song, J.-H., Kik, P. G. & Brongersma, M. L. Impact of substrates and quantum effects on exciton line shapes of 2D semiconductors at room temperature. *Nanophotonics* **12**, 3291–3300 (2023).

50. Atash Kahlon, A. *et al.* Importance of pure dephasing in the optical response of excitons in high-quality van der Waals heterostructures. *Phys. Rev. B* **112**, L041402 (2025).

51. Fang, H. H. *et al.* Control of the Exciton Radiative Lifetime in van der Waals Heterostructures. *Phys. Rev. Lett.* **123**, 067401 (2019).

52. Horng, J. *et al.* Engineering radiative coupling of excitons in 2D semiconductors. *Optica* **6**, 1443 (2019).

53. Meshulam, M. *et al.* Temperature-Dependent Optical and Polaritonic Properties of Excitons in hBN-Encapsulated Monolayer TMDs. *Adv. Opt. Mater.* e02535 (2025) doi:10.1002/adom.202502535.

54. Kim, H., Uddin, S. Z., Higashitarumizu, N., Rabani, E. & Javey, A. Inhibited nonradiative decay at all exciton densities in monolayer semiconductors. *Science* **373**, 448–452 (2021).

55. Brumme, T., Calandra, M. & Mauri, F. First-principles theory of field-effect doping in transition-metal dichalcogenides: Structural properties, electronic structure, Hall coefficient, and electrical conductivity. *Phys. Rev. B* **91**, 155436 (2015).

56. Ma, N. & Jena, D. Carrier statistics and quantum capacitance effects on mobility extraction in two-dimensional crystal semiconductor field-effect transistors. *2D Mater.* **2**, 015003 (2015).

57. Lien, D. H. *et al.* Electrical suppression of all nonradiative recombination pathways in monolayer semiconductors. *Science* **364**, 468–471 (2019).

58. Fan, S., Suh, W. & Joannopoulos, J. D. Temporal coupled-mode theory for the Fano resonance in optical resonators. *J. Opt. Soc. Am. A* **20**, 569 (2003).

59. Piper, J. R. & Fan, S. Total Absorption in a Graphene Monolayer in the Optical Regime by Critical Coupling with a Photonic Crystal Guided Resonance. *ACS Photonics* **1**, 347–353 (2014).

60. Liu, M. *et al.* A graphene-based broadband optical modulator. *Nature* **474**, 64–67 (2011).

61. Kuzmenko, A. B., Van Heumen, E., Carbone, F. & Van Der Marel, D. Universal optical conductance of graphite. *Phys. Rev. Lett.* **100**, 2–5 (2008).

62. Thureja, P. *et al.* Array-level inverse design of beam steering active metasurfaces. *ACS Nano* **14**, 15042–15055 (2020).

63. Das, S., Wang, Y., Dai, Y., Li, S. & Sun, Z. Ultrafast transient sub-bandgap absorption of monolayer $MoS_2$. *Light Sci. Appl.* **10**, 27 (2021).

64. Mohammadi Estakhri, N. & Alù, A. Wave-front Transformation with Gradient Metasurfaces. *Phys. Rev. X* **6**, 041008 (2016).

65. Shastri, K. & Monticone, F. Nonlocal flat optics. *Nat. Photonics* **17**, 36–47 (2023).

66. Overvig, A. C., Malek, S. C. & Yu, N. Multifunctional Nonlocal Metasurfaces. *Phys. Rev. Lett.* **125**, 017402 (2020).

67. Wang, W. *et al.* Clean assembly of van der Waals heterostructures using silicon nitride membranes. *Nat. Electron.* **6**, 981–990 (2023).

68. Kühner, L. *et al.* High-*Q* Nanophotonics over the Full Visible Spectrum Enabled by Hexagonal Boron Nitride Metasurfaces. *Adv. Mater.* **35**, 2209688 (2023).

69. Purdie, D. G. *et al.* Cleaning interfaces in layered materials heterostructures. *Nat. Commun.* **9**, 5387 (2018).

70. Xu, X. *et al.* Growth of 2D Materials at the Wafer Scale. *Adv. Mater.* **34**, 2108258 (2022).

71. Huang, Y. *et al.* Universal mechanical exfoliation of large-area 2D crystals. *Nat. Commun.* **11**, 2453 (2020).

72. Liu, F. *et al.* Disassembling 2D van der Waals crystals into macroscopic monolayers and reassembling into artificial lattices. *Science* **367**, 903–906 (2020).

73. Mannix, A. J. *et al.* Robotic four-dimensional pixel assembly of van der Waals solids. *Nat. Nanotechnol.* **17**, 361–366 (2022).

74. Dibos, A. M. *et al.* Electrically Tunable Exciton–Plasmon Coupling in a $WSe_2$ Monolayer Embedded in a Plasmonic Crystal Cavity. *Nano Lett.* **19**, 3543–3547 (2019).

75. Yuan, R., Lynch, J. & Jariwala, D. Ultra-compact plexcitonic electro-absorption modulator. *Device* **1**, 100002 (2023).

76. Miller, D. A. B. Attojoule Optoelectronics for Low-Energy Information Processing and Communications. *J. Light. Technol.* **35**, 346–396 (2017).

**For Table of Contents Only**

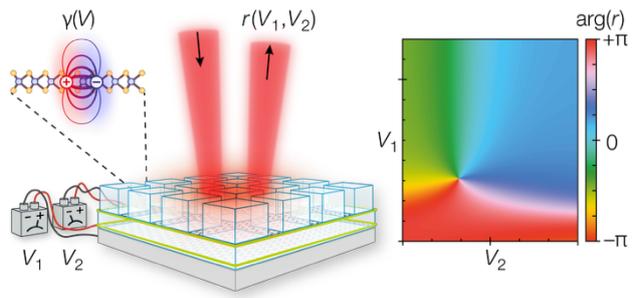